\documentclass[11pt,a4paper]{article}
\usepackage[utf8]{inputenc}
\usepackage{amsmath}
\usepackage{amsfonts}
\usepackage{amssymb}
\usepackage[left=2cm,right=2cm,top=2.5cm,bottom=2.5cm]{geometry}

\usepackage{slashed}
\usepackage{hyperref}
\usepackage{graphicx}
\usepackage{caption}
\usepackage{float}

\usepackage{cite}

\usepackage[labelformat=simple]{subcaption}

\usepackage{multirow}
\usepackage[dvipsnames]{xcolor}
\usepackage{multicol}
\usepackage{authblk}

\renewcommand{\theequation}{\arabic{section}.\arabic{equation}}

\usepackage{ifthen}
\usepackage{tikz}
\usepackage{xspace}
\usepackage{listings}
\usepackage[T1]{fontenc}

\definecolor{mediumjunglegreen}{rgb}{0.11, 0.21, 0.18}
\definecolor{bg}{HTML}{282828}

\newcommand*\cppin{\lstinline[language=c++]}

\newcommand*\pyin{\lstinline[language=c++]}

\lstnewenvironment{cpp}
{\lstset{language=c++}}
{}

\lstnewenvironment{py}
{\lstset{language=Python}}
{}

\lstnewenvironment{bash}
{\lstset{language=Python}}
{}

\lstnewenvironment{pseudo}
{\lstset{language=Python}}
{}

\newenvironment{run}[1]{``{\tt #1}"\xspace}

\newcommand{\ie}{{\em i.e.}\xspace}
\newcommand{\eg}{{\em e.g.}\xspace}
\newcommand{\GeV}{{\rm GeV}\xspace}

\newcommand{\MeV}{{\rm MeV}\xspace}

\def\nsc{{\tt NSC++}\xspace}
\def\mimes{{\tt MiMeS}\xspace}

\newcommand{\CPP}{{\tt C++}\xspace}

\newcommand{\PY}{{\tt python}\xspace}

\newcommand{\sR}{ s_{\rm R}\xspace}
\newcommand{\HR}{ H_{\rm R}\xspace}
\newcommand{\Hend}{ H_{\rm R,\,end}\xspace}
\newcommand{\rhoR}{ \rho_{\rm R}\xspace}
\newcommand{\rhoPhi}{ \rho_{\Phi}\xspace}
\newcommand{\GammaPhi}{ \Gamma_{\Phi}\xspace}

\newcommand{\fR}{ f_{\rm R}\xspace}
\newcommand{\fPhi}{f_{\Phi}\xspace}

\newcommand{\rhoRi}{ \rho_{\rm R\,i}\xspace}
\newcommand{\rhoPhii}{ \rho_{\Phi\,{\rm i}}\xspace}

\newcommand{\geff}{ g_{\rm eff}{}\xspace}
\newcommand{\heff}{ h_{\rm eff}{}\xspace}
\renewcommand{\dh}{ \delta_h{}\xspace}

\newcommand{\Ti}{ T_{\rm i}{}\xspace}

\newcommand{\ui}{ u_{\rm i}{}\xspace}

\newcommand{\ai}{ a_{\rm i}{}\xspace}

\newcommand{\Tend}{ T_{\rm end}{}\xspace}

\newcommand{\TEI}{ T_{\rm E1}{}\xspace}
\newcommand{\TEII}{ T_{\rm E2}{}\xspace}
\newcommand{\TDI}{ T_{\rm D1}{}\xspace}
\newcommand{\TDII}{ T_{\rm D2}{}\xspace}

\newcommand{\uEI}{ u_{\rm E1}{}\xspace}
\newcommand{\uEII}{ u_{\rm E2}{}\xspace}
\newcommand{\uDI}{ u_{\rm D1}{}\xspace}
\newcommand{\uDII}{ u_{\rm D2}{}\xspace}

\newcommand{\EI}{{\rm E1}{}\xspace}
\newcommand{\EII}{{\rm E2}{}\xspace}
\newcommand{\DI}{{\rm D1}{}\xspace}
\newcommand{\DII}{{\rm D2}{}\xspace}

\newcommand{\lrsb}[1]{\left[ #1 \right]}

\newcounter{NumArgs}

\newcommand{\eqs}[1]{\setcounter{NumArgs}{0}\foreach\i in{#1}{\stepcounter{NumArgs}}%
	\ifthenelse{\equal{\theNumArgs}{1}}{eq.~(\ref{#1})}%
	{\ifthenelse{\equal{\theNumArgs}{2}}%
		{eqs.~\foreach\i[count=\q]in{#1}{\ifthenelse{\equal{\q}{\theNumArgs}}{and (\ref{\i})}{(\ref{\i})~}}}%
		{eqs.~\foreach\i[count=\q]in{#1}{\ifthenelse{\equal{\q}{\theNumArgs}}{and (\ref{\i})}{(\ref{\i}),~}}}}}

\newcommand{\Eqs}[1]{\setcounter{NumArgs}{0}\foreach\i in{#1}{\stepcounter{NumArgs}}%
	\ifthenelse{\equal{\theNumArgs}{1}}{Eq.~(\ref{#1})}%
	{\ifthenelse{\equal{\theNumArgs}{2}}%
		{Eqs.~\foreach\i[count=\q]in{#1}{\ifthenelse{\equal{\q}{\theNumArgs}}{and (\ref{\i})}{(\ref{\i})~}}}%
		{Eqs.~\foreach\i[count=\q]in{#1}{\ifthenelse{\equal{\q}{\theNumArgs}}{and (\ref{\i})}{(\ref{\i}),~}}}}}

\newcommand{\refs}[1]{\setcounter{NumArgs}{0}\foreach\i in{#1}{\stepcounter{NumArgs}}%
	\ifthenelse{\equal{\theNumArgs}{1}}{(\ref{#1})}%
	{\ifthenelse{\equal{\theNumArgs}{2}}%
		{\foreach\i[count=\q]in{#1}{\ifthenelse{\equal{\q}{\theNumArgs}}{and (\ref{\i})}{(\ref{\i})~}}}%
		{\foreach\i[count=\q]in{#1}{\ifthenelse{\equal{\q}{\theNumArgs}}{and (\ref{\i})}{(\ref{\i}),~}}}}}

\newcommand{\Figs}[1]{\setcounter{NumArgs}{0}\foreach\i in{#1}{\stepcounter{NumArgs}}%
	\ifthenelse{\equal{\theNumArgs}{1}}{Figure~(\ref{#1})}%
	{\ifthenelse{\equal{\theNumArgs}{2}}%
		{Figures~\foreach\i[count=\q]in{#1}{\ifthenelse{\equal{\q}{\theNumArgs}}{and (\ref{\i})}{(\ref{\i})~}}}%
		{Figures~\foreach\i[count=\q]in{#1}{\ifthenelse{\equal{\q}{\theNumArgs}}{and (\ref{\i})}{(\ref{\i}),~}}}}}

\newcommand{\Gen}[2]{\setcounter{NumArgs}{0}\foreach\i in{#2}{\stepcounter{NumArgs}}%
	\ifthenelse{\equal{\theNumArgs}{1}}{#1.~(\ref{#2})}%
	{\ifthenelse{\equal{\theNumArgs}{2}}%
		{#1.~\foreach\i[count=\q]in{#2}{\ifthenelse{\equal{\q}{\theNumArgs}}{and (\ref{\i})}{(\ref{\i})~}}}%
		{#1.~\foreach\i[count=\q]in{#2}{\ifthenelse{\equal{\q}{\theNumArgs}}{and (\ref{\i})}{(\ref{\i}),~}}}}}

\author[ ]{Dimitrios Karamitros}
\affil[ ]{\em Department of Physics and Astronomy, The University of Manchester,}
\affil[ ]{\em Manchester M13 9PL, United Kingdom}

\affil[ ]{}
\affil[ ]{\textit{E-mail: } \href{mailto:dimitrios.karamitros@manchester.ac.uk}{\color{blue}{dimitrios.karamitros@manchester.ac.uk}}}

\title{\nsc: Non-Standard Cosmologies in {\tt C++}}
\begin{document}

\maketitle

\begin{abstract}
We introduce \nsc, a header-only \CPP library that simulates the evolution of the plasma and a decaying fluid in the early Universe.
\nsc can be used in \CPP programs or called directly from \PY scripts without significant overhead.
There is no special installation process or external dependencies. Furthermore, there are example programs that can be modified to handle several cases.\\ 
\end{abstract}


\noindent
{\bf Program summary:}\\

{\sl 
	Program title: \nsc.\\
	
	Developer's respository link: \href{https://github.com/dkaramit/NSCpp}{https://github.com/dkaramit/NSCpp}.\\
	
	Programming language: \CPP ({\tt C++17}$+$) and \PY ({\tt 3.7}$+$).\\
	
	Licensing provisions: MIT license.\\
	
	Nature of problem: Solves equations that describe the evolution of the plasma along with a fluid that increases the entropy of the plasma during the early Universe.\\
	
	Solution method: Embedded Runge-Kutta for the numerical integration of the system of differential equations. The user can choose between explicit and Rosenbrock methods. There are several of Butcher tableaux already implemented, but the user can implement their own. The interpolations of the relativistic degrees of freedom are accomplished by cubic spline interpolation.\\ 
	
	Restrictions: The value of the pressure over the energy density and the energy loss parameter of the fluid are assumed to be constant. It is also assumed that there is no energy leaking from the plasma to the fluid.
}

\tableofcontents

\section{Introduction}\label{sec:intro}
\setcounter{equation}{0}
The configuration of the components of the Universe today might have been influenced by its nature at times that cannot be probed. Therefore, there is the possibility that the Universe experienced a period of non-standard evolution. A scenario that deviates from the standard cosmological case is called non-standard cosmology (NSC). 
NSCs in general affect the evolution of other components of the Universe, and their detailed examination can help us probe the nature of the Universe at temperatures inaccessible by direct observations (\eg of CMB). For example, the study of dark matter within NSCs can allow possibilities that are otherwise excluded. This means that if dark matter is observed and turns out to be excluded under the standard cosmological model, then its nature can reveal aspects of the Universe at times relevant to the dark matter production (see,~\eg~\cite{McDonald:1989jd,DEramo:2017gpl,Redmond:2017tja,DEramo:2017ecx,Bernal:2020bfj,Arias:2020qty,Arias:2021rer,Barman:2021ifu,Dienes:2021woi,Banerjee:2022fiw,Hardy:2018bph,Bernal:2018kcw,Arias:2019uol,Allahverdi:2019jsc,Bernal:2019mhf,Cosme:2020mck}). 
Several non-standard cosmological scenarios have been studied in the literature (\eg refs.~\cite{Vilenkin:1982wt,Coughlan:1983ci,Ratra:1987rm,Giudice:2000ex,Gardner:2004in,Dalianis:2018afb} with some reviews found in refs.~\cite{Tsujikawa:2013fta,Allahverdi:2020bys}), however there is no available tool that will automate and standardise the simulation the evolution of the Universe. \nsc is an attempt to accomplish this by being an easy-to-use and modifiable \CPP library that can handle the most common NSCs. 
Since in most cases the expansion mode of the Universe is just a background, \nsc can be used to simulate it and use the results to study the evolution of other components of the Universe (\eg dark matter or lepton number) under this background. 
In particular, \nsc can be used along with \mimes~\cite{Karamitros:2021nxi}, since it can output the evolution of the temperature and Hubble parameter which can be used as an input by \mimes.

We note that there is another package, built specifically for the evolution of cosmic relics in matter dominated Universe, {\tt EvoEMD}~\cite{Dutra:2021phm}. \nsc is different because the evolution of the temperature of the plasma is computed by manipulating the equation that describes the evolution of the entropy density, while {\tt EvoEMD} solves
\begin{equation}
	\dfrac{d\rhoR}{dt} = -4H\rhoR + C_{\rm EMD} \;,
	\label{eq:EvoMD}
\end{equation}
with  $C_{\rm EMD}$ the additional matter contribution. This equation relies on the approximation $\rhoR \approx 3/4 \, sT$ that does not hold in general. Also, \nsc allows the user to consider different equations of state for the fluid, $\Phi$. 
As we explain in this article, \nsc adheres to design principles similar to \mimes~\cite{Karamitros:2021nxi}; \ie the general usage of both is almost identical, \nsc only needs a \CPP compiler that supports the {\tt c++17} standard with no other external dependences, and there will always be dedicated versions of the underline numerical libraries~\cite{NaBBODES,SimpleSplines} guaranteed to work with the current version of \nsc. Moreover, there are multiple Runge-Kutta (RK) methods~\cite{NaBBODES} included with the possibility to implement new ones by the user. This can help the users inspect the numerical validity of the results.

This article is organised as follows. Section~\ref{sec:equations} is brief reminder of the system that \nsc simulates and a description of the internal notation it uses. The next section describes the usage of \nsc in general, from downloading to compiling, along with a brief description of the various classes and functions available to the user. In Section~\ref{sec:example} there are complete examples written in both \CPP and \PY, and a brief mention of all the results one can output. Section~\ref{sec:summary} summarises the article and provides future directions.
Appendices~\refs{app:cpp,app:py} describe in detail the implementation of the classes available to the user, while in Appendix~\refs{app:usr_input} there are tables that summarise all the possible inputs and options. 

\section{Evolution equations}\label{sec:equations}
\setcounter{equation}{0}

In order to model an NSC, we assume that the Universe at early times was dominated by two components; the plasma and a fluid ($\Phi$) with an equation of state
\begin{equation}
	p_{\Phi} = (c/3-1) \rhoPhi \;,
	\label{eq:EOS}
\end{equation}  
with $c$ a constant. 

The plasma is assumed to be always in thermal equilibrium, with energy and entropy densities defined as
\begin{align}
	\sR(T)&=\dfrac{2\pi^2}{45} \,  \heff(T) \, T^3 \label{eq:s_def}\\
	\rhoR(T)&=\dfrac{\pi^2}{30} \,  \geff(T) \, T^4 \label{eq:rho_def} \;,
\end{align}
where $\heff(T)$ and $\geff(T)$ count the effective relativistic degrees of freedom (RDOF) of the plasma.

If $\Phi$ dominates for some temperature range, it must lose energy to the plasma, since the contribution to the energy budget of the Universe of such fluid is severely constraint at temperatures lower than $T \sim \mathcal{O}(10)~\MeV$~\cite{Allahverdi:2020bys}. We model this by introducing a constant energy loss rate, $\GammaPhi$. The equations that describe the evolution of both components are
\begin{align}
	\dfrac{d\sR(T)}{dt} &= -3 \, H(T)  \, \sR(T) + \dfrac{\GammaPhi}{T} \, \rhoPhi(T) \label{eq:dsdt} \\ 
	\dfrac{d\rhoPhi(T)}{dt} &= - c \, H(T) \, \rhoPhi(T) - \GammaPhi \, \rhoPhi(T) \label{eq:drhoPhidt} \;,
\end{align}
with $H$ the Hubble parameter given by
\begin{equation}
	H(T) = \sqrt{\dfrac{8}{3\,m_P^2} \ \lrsb{\rhoR(T) + \rhoPhi(T)}}\;,
	\label{eq:H_def}
\end{equation}
where $m_P = 1.22 \times 10^{19}~\GeV$.

Following refs.~\cite{Bernal:2019mhf,Arias:2020qty}, we can parametrise $\GammaPhi$ in terms of the temperature that $\Phi$ would decay ($\Tend$) if the Universe was radiation dominated, as
\begin{equation}
	\GammaPhi \equiv \HR(T=\Tend)\;,
	\label{eq:Tend_def}
\end{equation}
where $\HR$ the Hubble parameter for a radiation dominated Universe obtained from \eqs{eq:H_def} with $\rhoPhi=0$.

\subsection{Notation}\label{sec:notation}

In order to solve this system of equations, we change the integration variable to 
\begin{equation}
	u=\log\frac{a}{\ai}\;,	
	\label{eq:u_def}
\end{equation}
with $a$ the scale factor of the Universe, and $\ai$ its value at some initial time. The initial condition can now be chosen at $\ui=0$, corresponding to $T=\Ti$ and an initial value for $\rhoPhi=\rhoPhii$. The form of \eqs{eq:u_def} implies that the evolution of the Universe depends only on its relative scale; \ie if $\ai \neq 0$, its value does not matter.

We can reparametrise the temperature of the plasma and the energy density of $\Phi$ as 
\begin{align}
	T(u) &= \Ti  \, \fR(u) \, e^{-u}  \label{eq:fR_def} \\ 
	\rhoPhi(u) &= \rhoPhii \, \fPhi(u) \label{eq:fPhi_def} \, e^{-cu}\;,
\end{align}
with $\fR(u)$ and $\fPhi(u)$ functions of $u$ that obey the initial conditions $\fR(u=0)=\fPhi(u=0)=1$ and the equations
\begin{align}
	\dfrac{d\log \fR }{du} &=  1- 1/\dh- \dfrac{1}{3}\dfrac{\GammaPhi}{H\,\dh} \, \dfrac{\rhoPhi}{T\,s}   \label{eq:dlogfRdu} \\ 
	\dfrac{d\log\fPhi}{du} &= -\dfrac{\GammaPhi}{H} \label{eq:dlogfPhidu} \;,
\end{align}
with 
\begin{equation}
	\dh = 1+ \dfrac{1}{3} \dfrac{d\log\heff}{d\log T}\;.
	\label{eq:dh_def}
\end{equation}
Notice that for $\GammaPhi=0$ and $\dh=1$, the functions introduced are $\fR=\fPhi=1$. These are the equations that \nsc actually solves, using as user input for $\Tend$, $c$, $\Ti$, and $r=\rhoRi/\rhoPhii$. Then, it automatically transforms them to the temperature of the plasma and the fluid energy density. From these, the user can obtain all other relevant cosmological quantities such as the Hubble parameter or the entropy density of the plasma.

\nsc closely follows the notation of ref.~\cite{Arias:2020qty}, identifying (at most) four points where the behaviour of the system changes; $\EI$, $\DI$, $\EII$, and $\DII$. The point $\EI$ ($\EII$) is defined when $\rhoPhi=\rhoR$ for the first (second) time. The point $\DI$ ($\DII$) is defined as the point at which $\lrsb{ (\Hend/H) \, (\rhoPhi/\rhoR) }$ exceeds (drops below) $40\%$; \ie before $\DI$ and after $\DII$ the energy loss rate from the fluid to the plasma is less than $10\%$ of the free dilution rate of $\rhoR$.
For the sake of performance, \nsc identifies these points approximately during the integration of~\eqs{eq:dlogfRdu,eq:dlogfPhidu}. 

\begin{center}
	\begin{figure}[t!]
		\begin{subfigure}{0.5\textwidth}
			\includegraphics[width=1\textwidth]{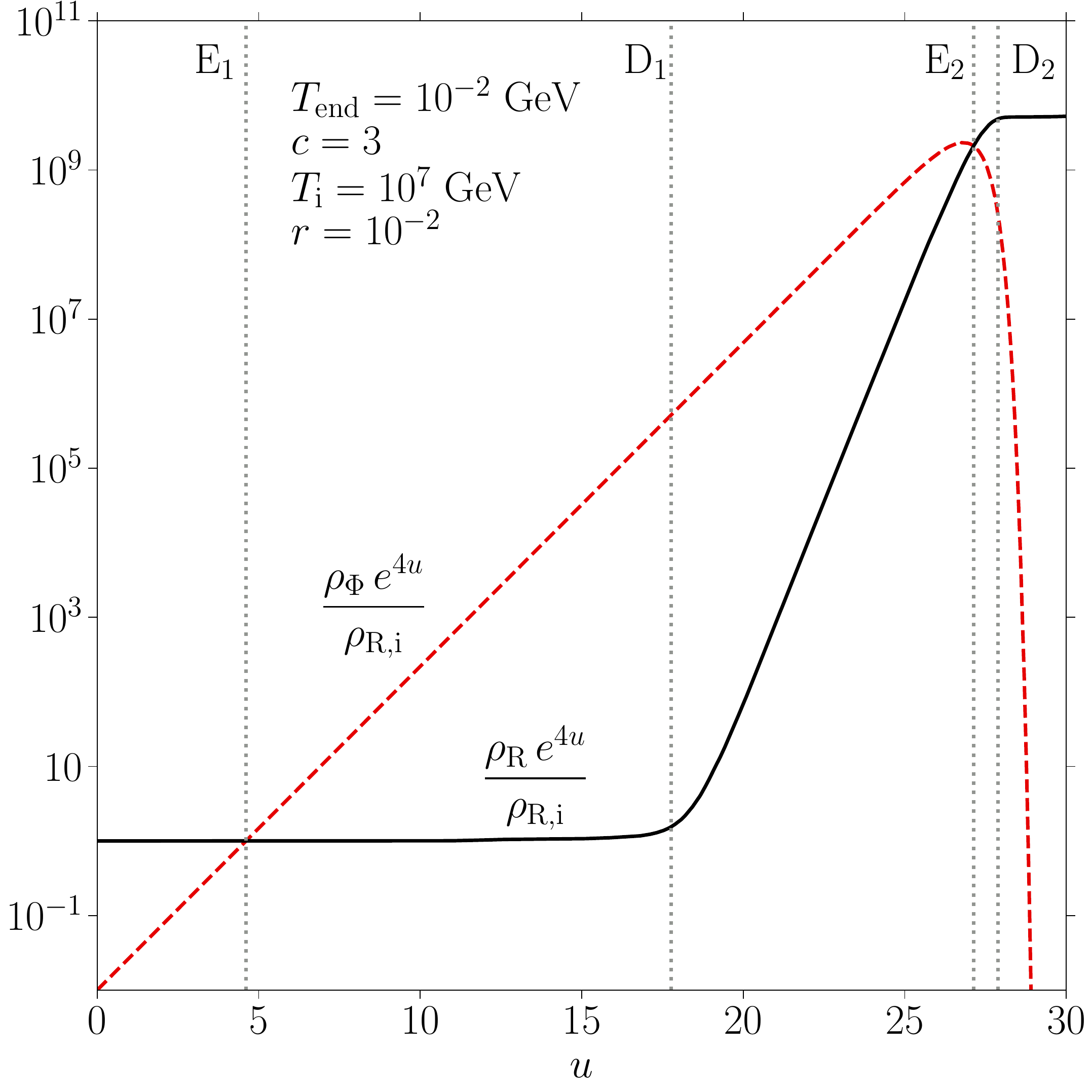}
			\caption{}
			\label{fig:EMD}
		\end{subfigure}
		\begin{subfigure}{0.5\textwidth}
			\includegraphics[width=1\textwidth]{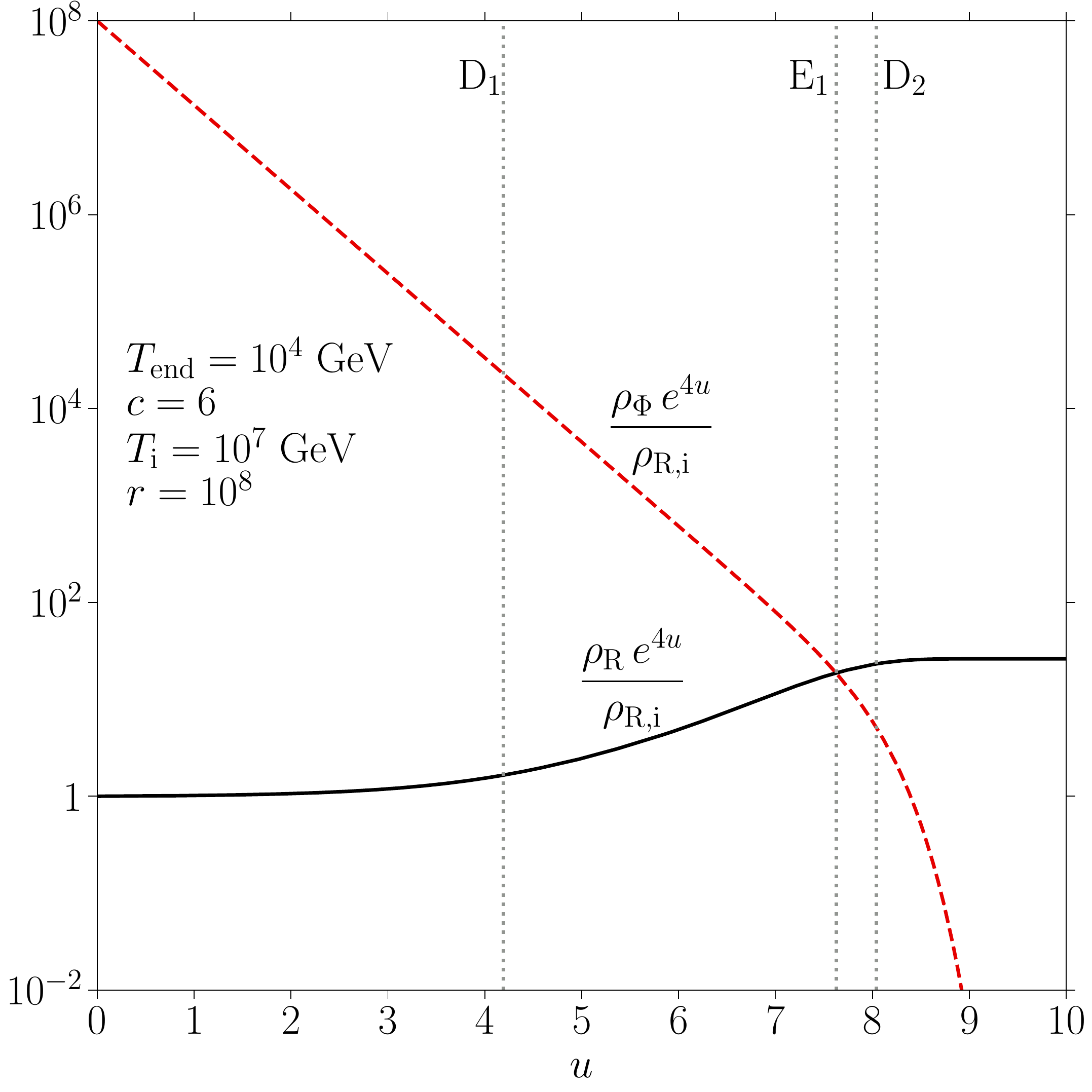}
			\caption{}
			\label{fig:EKD}
		\end{subfigure}
		\caption{
			({\bfseries a}) The evolution of the comoving energy densities of radiation (black) and $\Phi$ (dashed red) in the case of early matter domination ($c=3$) with $\Tend=10^{-2}~\GeV$, $\Ti=10^7~\GeV$, and $r=10^{-2}$. 
			({\bfseries b}) The same for kination ($c=6$) with $\Tend=10^{4}~\GeV$, $\Ti=10^7~\GeV$, and $r=10^{8}$.
			The grey vertical lines show the approximate location of the points ${\rm E_{1,2}}$ and ${\rm D_{1,2}}$.
		} 
		\label{fig:evolution_examples}
	\end{figure}
\end{center}

In general, there are two cases that exhibit qualitatively different evolution, which are shown in \Figs{fig:evolution_examples}. In \Figs{fig:EMD}, we show the evolution of the comoving energy densities of radiation and $\Phi$ as functions of $u$ for $\Tend=10^{-2}~\GeV$, $c=3$, $\Ti=10^7~\GeV$, and $r=10^{-2}$. In \Figs{fig:EKD}, we show the evolution of these quantities for $\Tend=10^{4}~\GeV$, $c=6$, $\Ti=10^7~\GeV$, and $r=10^{8}$. 
These two figures correspond to the two qualitative typical behaviours we expect for $c<4$ and $c>4$. In the former, $\Phi$ increases its energy contribution, until it decays away. In the latter,  $\rhoPhi$ monotonically decreases. In this case, the Universe will become radiation dominated regardless $\GammaPhi$, but if $\rhoPhii$ and $\GammaPhi$ are large enough the energy of the plasma will increase.  Notice that for $c>4$, There can only be one point of equality between $\rhoR$ and $\rhoPhi$, which we call $\EI$ as this is the naming conversion in \nsc.

\section{\nsc usage}\label{sec:first_steps}
\setcounter{equation}{0}
\nsc can be obtained directly form github. The safest choice is to download the ``stable'' branch, which will always be the latest tested version. The preferred method for downloading \nsc is by running:~\footnote{The {\tt git} utility can be installed by following \href{https://github.com/git-guides/install-git}{https://github.com/git-guides/install-git}.}
\begin{lstlisting}
	git clone -b stable https://github.com/dkaramit/NSCpp
\end{lstlisting}

\nsc relies on {\tt NaBBODES}~\cite{NaBBODES} and {\tt SimpleSplines}~\cite{SimpleSplines}, which are developed independently. In order to get \nsc with the latest version of these libraries, one needs to run the following commands
\begin{lstlisting}
	git clone https://github.com/dkaramit/NSCpp.git
	cd NSCpp
	git submodule init
	git submodule update --remote
\end{lstlisting}
These commands download the {\tt master} branch of \nsc, with the latest versions of {\tt NaBBODES} and {\tt SimpleSplines}. All versions that are not labelled as stable must be treated with caution, as they may not work properly. 

By executing \run{bash configure.sh} in the root directory of \nsc, the library will be available for usage within a \CPP program. This {\tt bash} script, writes some paths in some files for convenience, formats the data for the RDOF in an acceptable format (if a {\em relative} path is given in the {\tt NSCpp/Paths.mk file}), and creates some directories that may be useful later. In order to call \nsc inside \PY, \run{make lib} must also be run,  which will compile shared libraries that needed for the \PY interface.

\nsc already comes with several examples in \CPP and \PY, which can be found in the directory {\tt NSCpp/UserSpace}. In order to compile the \CPP examples, one just has to run \run{make examples} in the root directory of \nsc or {\tt make} inside the directory of each \CPP program.

There is also an option to compile some test programs using {\tt make check}, which will create executables inside {\tt exec}. These can then be executed, in order to ensure that the code actually works; \eg no segmentation fault occurs. 

\subsection{Assumptions}
\nsc is designed to be as model agnostic as possible. It can only be used for constant $c$ and $\GammaPhi$, and in cases where there are no plasma energy leaks into the fluid. In cases that these assumptions do not hold, knowledge of the exact nature of $\Phi$ is generally needed.  It is also assumed that the plasma does not fall out of thermal equilibrium between the temperatures we integrate (defined later as {\tt Ti} and {\tt TSTOP}), and its energy and entropy densities are always given by \eqs{eq:s_def,eq:rho_def}.

\subsection{Using \nsc}\label{sec:Cpp_usage} 
Out-of-the-box, and after \run{bash configure.sh} is run successfully, \nsc can be used in a \CPP program by including in the source file the following line:
\begin{cpp}
	#include "NSCpp.hpp"
\end{cpp}
Notice that if this file is not in the root directory of \nsc, we need to compile it using the flag {\tt -Ipath-to-root}, "path-to-root" the relative (or absolute) path to the root directory of \nsc; \eg if the {\tt .cpp} is in the {\tt NSCpp/UserSpace/Cpp/NSC} directory, this flag should be {\tt -I../../../}. 
 
The class that deals with the system~\refs{eq:dlogfRdu,eq:dlogfPhidu}, is the \cppin{nsc::Evolution<LD,SOLVER,METHOD>}. In order to begin, we declare an instance of this class as
\begin{cpp}
    nsc::Evolution<LD,Solver,Method> BE;
\end{cpp}
Here, \cppin{LD} should be the numeric type to be used. Only floating point types are allowed, and it is recommended to use \cppin{double} or \cppin{long double}. Moreover \cppin{Solver} and \cppin{Method} depend on the type of RK method the user chooses. The available choices are shown in Table~\ref{tab:template-arguments}. 

The evolution of the Universe depends on the evolution of the plasma, which is encoded in the class \cppin{nsc::Cosmo<LD>}. An instance of this class is declared as 
\begin{cpp}
	nsc::Cosmo<LD> plasma('path_to_data', minT, maxT);
\end{cpp}
Here, \cppin{'path_to_data'} is the (relative of absolute) path to a file with tabulated data  of the relativistic degrees of freedom of the plasma . It is assumed that the file has three columns with: $T$ in $\GeV$, $\heff$, and $\geff$. The temperature given in this file has to increase monotonically, otherwise the program exits with error core $1$.~\footnote{\nsc already includes the standard model RDOF found in~\cite{Saikawa:2020swg} in the directory {\tt NSCpp/src/data}.} If a path is provided in the {\tt NSCpp/Paths.mk} file before \run{bash configure.sh} is run, there will be a global (constant) variable called \cppin{cosmo_PATH} which can be used after  {\tt NSCpp.hpp} is included.
The arguments \pyin{minT} and \pyin{maxT} dictate the temperatures between which $\geff$ and $\heff$ will be interpolated. For temperatures beyond these, both  $\geff$ and $\heff$  are assumed to be constant. A detailed description of this class can be found in the Appendix of ref.~\cite{Karamitros:2021nxi}.

The system can be solved by including:
\begin{cpp}
	BE.solveNSC(TEND, c, Ti, ratio, TSTOP, umax, &plasma,
				 	 {
					 .initial_step_size=initial_step_size, .minimum_step_size=minimum_step_size,
					 .maximum_step_size=maximum_step_size, .absolute_tolerance=absolute_tolerance, 
					 .relative_tolerance=relative_tolerance, .beta=beta, 
					 .fac_max=fac_max, .fac_min=fac_min, .maximum_No_steps=maximum_No_steps
					 }
					);
\end{cpp}
The various parameters are:
\begin{enumerate}
	\item {\tt TEND}: the values of the $\Tend$ (in $\GeV$) as introduced in \eqs{eq:dsdt}.~\footnote{Note that $\GammaPhi=0$ corresponds to $\Tend=0$.}
	
	\item {\tt c}: the parameter that characterises the equation of state of the fluid, defined as in \eqs{eq:EOS}.
	
	\item {\tt Ti}: the initial temperature (in $\GeV$) of integration. Has to be above {\tt TEND}.
	
	\item {\tt ratio}: the value of $\rhoPhi/\rhoR$ at {\tt Ti}. 
	
	\item {\tt TSTOP}: If the temperature drops below this, integration stops. 
	
	\item {\tt umax }: If $u>${\tt umax} the integration stops. It can be used to stop internation if relevant bounds (\eg~\cite{Planck:2018jri}) are violated. It is also a failsafe parameter that ensures that integration stops even in cases where the Universe expands extremely fast.
	
	\item {\tt plasma }: A pointer to an instance of the \cppin{Cosmo} class.
	
	\item {\tt initial\_stepsize} (optional): Size of the initial step the solver takes. 
	
	\item {\tt minimum\_stepsize} (optional): The step-size will be limited to be above number. 
	
	\item {\tt maximum\_stepsize} (optional): The step-size will be limited to be below number. 
	
	\item {\tt absolute\_tolerance} (optional): Absolute tolerance of the RK solver.
	
	\item {\tt relative\_tolerance} (optional): Relative tolerance of the RK solver.
	
	\item {\tt beta} (optional): Aggressiveness of the adaptation process of the solver. Generally, it should be around but less than 1.
	
	\item {\tt fac\_max},  {\tt fac\_min} (optional): The solver will keep step-size to from increasing more than fac\_max and less than fac\_min at each iteration. This makes adaptation of the step-size more stable.
	
	\item {\tt maximum\_No\_steps} (optional): Maximum steps the solver can take. Quits if this number is reached even if the temperature is still larger than {\tt TEND}. 
\end{enumerate}
A detailed description of the  effect of the various optional parameters on the operation of the ODE solver can be found in the Appendices of ref.~\cite{Karamitros:2021nxi}. Notice that the optional parameters are passed using aggregation.~\footnote{This means that all arguments inside {\tt \{...\}} define a temporary an instance of a \cppin{struct} that simply holds these arguments.}

The \cppin{solveNSC} function returns \cppin{true} if the temperature reached \cppin{TSTOP} for $u<=$\cppin{umax} and \cppin{false} otherwise. This can be used to test whether the integration completed successfully.
The evolution of $T$ and $\rhoPhi$ is stored in \cppin{BE.T} and \cppin{BE.rhoPhi}, with each element corresponding to integration points $u$ (accessed as \cppin{BE.u}). The local errors of $T$ and $\rhoPhi$ correspond to the variables \cppin{BE.dT} and \cppin{BE.drhoPhi}. 
One can also obtain the points {\tt BE.TE1} ($\TEI$), {\tt BE.TE2} ($\TEII$), {\tt BE.TD1} ($\TDI$), {\tt BE.TD2} ($\TDII$). The corresponding values of $u$ are {\tt BE.uE1}, {\tt BE.uE2}, {\tt BE.uD1}, {\tt BE.uD2}.

\subsection{Using \nsc in \PY}\label{sec:py_usage}
The \PY interface modules are in the directory {\tt NSCpp/src/interfacePy}. The usage of the classes is made as similar as possible to the \CPP case. However, one should keep in mind that the various template arguments discussed in the \CPP case have to be chosen at compile-time. That is, for the \PY interface, one needs to choose the numeric type and RK method to be used when the shared libraries are compiled. This is accomplished by assigning the relevant variables in {\tt NSCpp/Definitions.mk} before running \run{bash configure.sh} and \run{make lib}. The various options are the same as in Section 4 of ref.~\cite{Karamitros:2021nxi}, and summarised in Table~\ref{tab:compile_time-options}.

The two relevant classes are defined in the module {\tt interfacePy}, and can be imported in a \PY script as 
\begin{py}
	from sys import path as sysPath
	sysPath.append('path_to_src')
	from interfacePy.Evolution import Evolution
	from interfacePy.Cosmo import Cosmo
\end{py}
The string {\tt 'path\_to\_src'} must be the relative path to the {\tt NSCpp/src} directory. For example, if the script is located in {\tt NSCpp/UserSpace/Python}, {\tt 'path\_to\_src'} should be {\tt '../../src'}.

We can define an {\tt Evolution} instance as follows 
\begin{py}
	BE=Evolution()
\end{py}

An instance of the \pyin{Cosmo} class is declared as 
\begin{cpp}
	plasma=Cosmo('path_to_data', minT, maxT)
\end{cpp}
The arguments are the same as in \CPP.

The system, then, is solved by running:
\begin{cpp}
	BE.solveNSC(TEND, c, Ti, ratio, TSTOP, umax, plasma,
						initial_step_size, minimum_step_size, maximum_step_size, absolute_tolerance, 
						relative_tolerance, beta, fac_max, fac_min, maximum_No_steps)
					  )
\end{cpp}
The arguments are identical to the \CPP case, outlined in Table~\ref{tab:solveNSC-input}, but the instance of \pyin{Cosmo} (\pyin{plasma}) is passed by value. A brief description of this function can also be found by running \pyin{?BE.solveNSC} after loading the module. This function returns \cppin{True} (\cppin{False}) if the temperature reached \cppin{TSTOP} for $u<=$\cppin{umax} ($u>$\cppin{umax}), and the time it took to execute it in seconds.
In contrast to the \CPP implementation, this only gives us access to the points where the behaviour changes; the corresponding variables are {\tt BE.TE1}, {\tt BE.TE2}, {\tt BE.TD1}, {\tt BE.TD2}, {\tt BE.uE1}, {\tt BE.uE2}, {\tt BE.uD1}, and {\tt BE.uD2}. In order to get the evolution of $T$ and $\rhoPhi$, we need to run 
\begin{py}
	BE.getPoints()
\end{py}
This will fill the \pyin{numpy}~\cite{harris2020array} arrays with the integration variable and the solutions; \pyin{BE.u}, \pyin{BE.T}, and \pyin{BE.rhoPhi}. The following line:
\begin{py}
	BE.getErrors()
\end{py}
gives access to the local errors \pyin{BE.dT} and \pyin{BE.drhoPhi}.

It is essential to manually delete both \pyin{BE} and \pyin{plasma} if they are no longer needed or before reassigning them. The reason is that there are underlying {\tt C}-pointers that manage the \PY interface for each instance of the classes. If the instances are reassigned without deletion, these pointers can no longer be accessed. This is a memory-leak, which can only be fixed by exiting the script. 
That is, one must simply run 
\begin{py}
	del BE, plasma
\end{py}  
after {\tt BE} and {\tt plasma} served their purpose or before reassigning these variables.

\section{Example}\label{sec:example}
\setcounter{equation}{0}
In this section we will show an example code that solves the system of \Figs{fig:EMD} in both \CPP and \PY.
\subsection{\CPP}
In \CPP, we need to include the header file \cppin{NSCpp.hpp} from the root directory of \nsc. Then, we declare an instance of the \cppin{Evolution} class. The system is solved by calling the \cppin{solveNSC} function with the inputs described in Table~\ref{tab:solveNSC-input}. The following code can be used to solve the system~\refs{eq:dsdt,eq:drhoPhidt}. This code will print the values of $T_{\rm E_{1,2}}$ and $T_{\rm D_{1,2}}$, and all the integration points with their local errors.
\begin{cpp}
#include<iostream>
#include<iomanip>

//Include everything you need from NSC++
#include"NSCpp.hpp"

int main(){
	//Use cosmo_PATH to interpolate ?$\heff$? and ?$\geff$? from T=0 to T=mP.
	nsc::Cosmo<long double> plasma(cosmo_PATH, 0, nsc::Cosmo<long double>::mP);
	
	//Declare Evolution instance using the Rosenbrock method RODASPR2
	nsc::Evolution<long double,1,RODASPR2<long double>> BE;
	
	//Declare parameters
	long double TEND=1e-2, c=3, Ti=1e7, ratio=1e-2, TSTOP=1e-4, umax=200;
	
	bool check=BE.solveNSC(TEND, c, Ti, ratio, TSTOP, umax, &plasma,
				{
					.initial_step_size=1e-2, .minimum_step_size=1e-8, .maximum_step_size=1e-2, 
					.absolute_tolerance=1e-11, .relative_tolerance=1e-11, .beta=0.9, 
					.fac_max=1.2, .fac_min=0.8, .maximum_No_steps=10000000
				});
	
	if(check){
		// If the solver returns true, the solver probably worked.
		std::cout<<std::setprecision(5);
		std::cout<<BE.TE1<<"\t"<<BE.TE2<<"\t"<<BE.TD1<<"\t"<<BE.TD2<<"\n";
		
		//print the results and the errors
		for(size_t i=0; i<BE.pointSize; ++i ){
			std::cout<<std::left<<std::setw(15)<<BE.u[i]<<std::setw(15);
			std::cout<<std::left<<BE.T[i]<<std::setw(15)<<BE.dT[i]<<std::setw(15);
			std::cout<<std::left<<BE.rhoPhi[i]<<std::setw(15)<<BE.drhoPhi[i]<<"\n";
		}
	}else{
		// If the solver returns false, you may need a larger umax.
		std::cerr<<"Something went wrong. Try using larger value for umax\n";
		exit(1);
	}
	
	return 0;
}
\end{cpp}
Notice that if \cppin{solveNSC} returns \cppin{false}, there is an error message printed, and the code exits  with error code $1$.

In this program we have used \cppin{long double} as the numeric type in all declarations, with the {\tt RODASPR2} Rosenbrock method~\cite{RangAngermann2005} for the RK solver. Assuming that this code is written in a file with name {\tt example.cpp}, we can compile it with 
\begin{bash}
	g++ -std=c++17 -O3 -lm -I'path_to_root' example.cpp -o example.run	
\end{bash}
or
\begin{bash}
	clang -lstdc++ -std=c++17 -O3 -lm -I'path_to_root' example.cpp -o example.run	
\end{bash}
with \cppin{'path_to_root'} the root directory of \nsc. This will create the executable {\tt example.run}. Note that the variable {\tt cosmo\_PATH} is defined in {\tt NSCpp/src/misc\_dir/path.hpp}, created by the script {\tt NSCpp/configure.sh} if a {\em relative} path is given in {\tt NSCpp/PATHS.mk}. As This variable is the absolute path to the RDOF file, the executable can be copied to any other place in the same system.

\subsection{\PY}
The \PY interface of \nsc can handle same system as in the previous subsection. This is shown in the following code:
\begin{py}
# append the path to the src directory 
from sys import path as sysPath
sysPath.append('../src')

#load the NSC module
from interfacePy.Evolution import Evolution 

# load Cosmo and the Planck mass
from interfacePy.Cosmo import Cosmo,mP 

#This gives you access to the path of the rdof file.
from misc_dir.path import cosmo_PATH

#Instance of the Cosmo class. Interpolate from T=0 to T=mP
plasma=Cosmo(cosmo_PATH,0,mP)

# Evolution instance
BE=Evolution()

# solve the system
check,time=BE.solveNSC(TEND=1e-2, c=3, Ti=1e7, ratio=1e-2, umax=500, TSTOP=1e-4, plasma=plasma,
				initial_step_size=1e-2, minimum_step_size=1e-8, maximum_step_size=1e-2, 
				absolute_tolerance=1e-11, relative_tolerance=1e-11, beta=0.9, fac_max=1.2, 
				fac_min=0.8,maximum_No_steps=10000000)

if check:
	# get points 
	BE.getPoints()
	# get errors
	BE.getErrors()

print(BE.TE1,BE.TE2,BE.TD1,BE.TD2)

for i,u in enumerate(BE.u):
	print(u,BE.T[i],BE.dT[i],BE.rhoPhi[i],BE.drhoPhi[i])
else:
	print("Something went wrong. Try using larger value for umax")
	exit(1)

#run the destructors
del BE
del plasma
\end{py}
This script is assumed to be located in a subdirectory of the root directory of \nsc, \eg inside {\tt NSCpp/UserSpace}. If it is in another directory, then only the third line should be changed to \pyin{sysPath.append('path_to_root')}, with \pyin{'path_to_root'} the path to the root directory of \nsc.

\subsection{Result} 
In both \CPP and \PY, the results we obtain are identical, and evolution of the comoving energy densities is given by \Figs{fig:EMD}. 
\begin{figure}[t]
	\begin{subfigure}{0.5\textwidth}
		\includegraphics[width=1\textwidth]{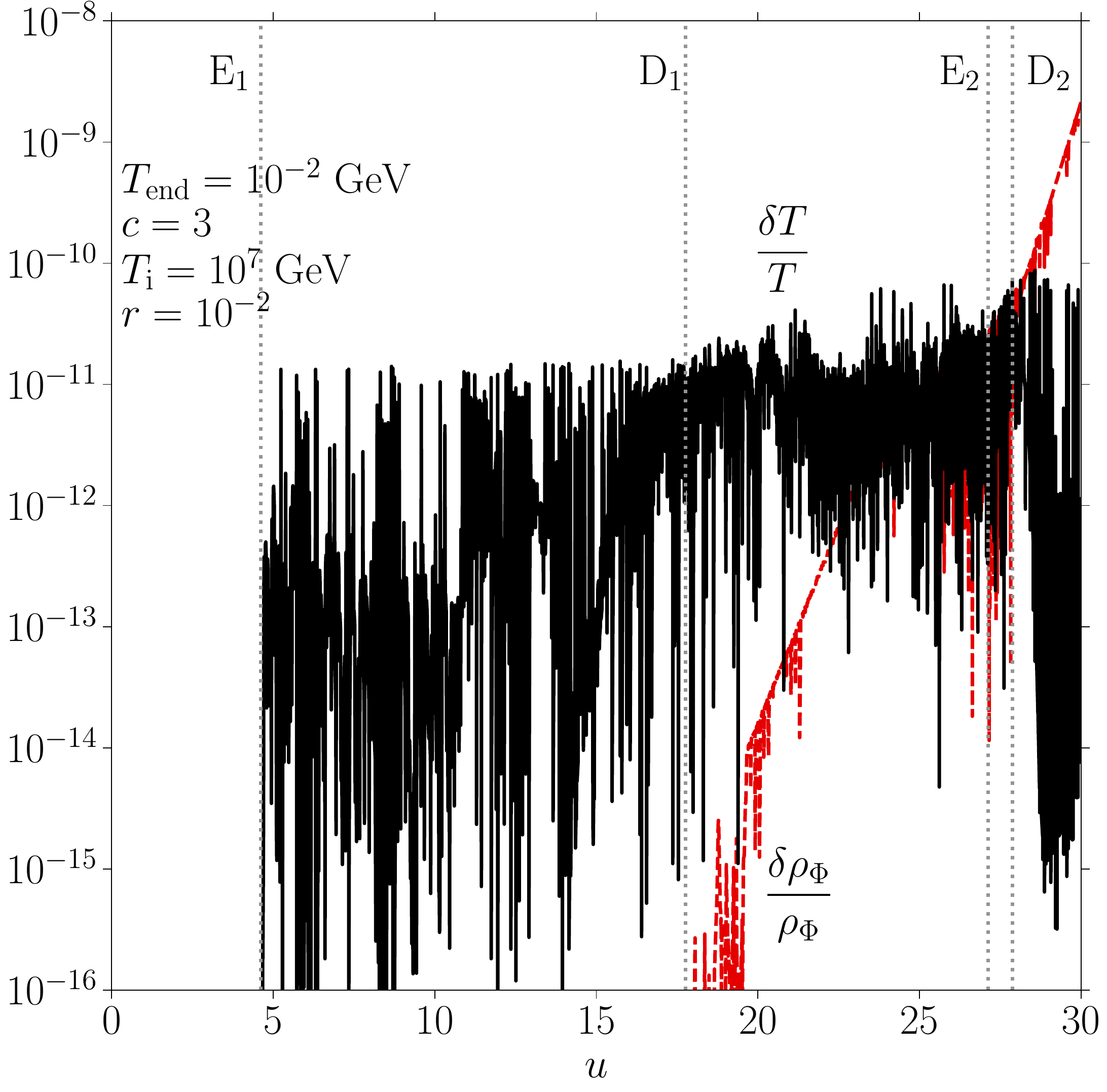}
		\caption{}
		\label{fig:EMD_err}
	\end{subfigure}
	\begin{subfigure}{0.5\textwidth}
		\includegraphics[width=1\textwidth]{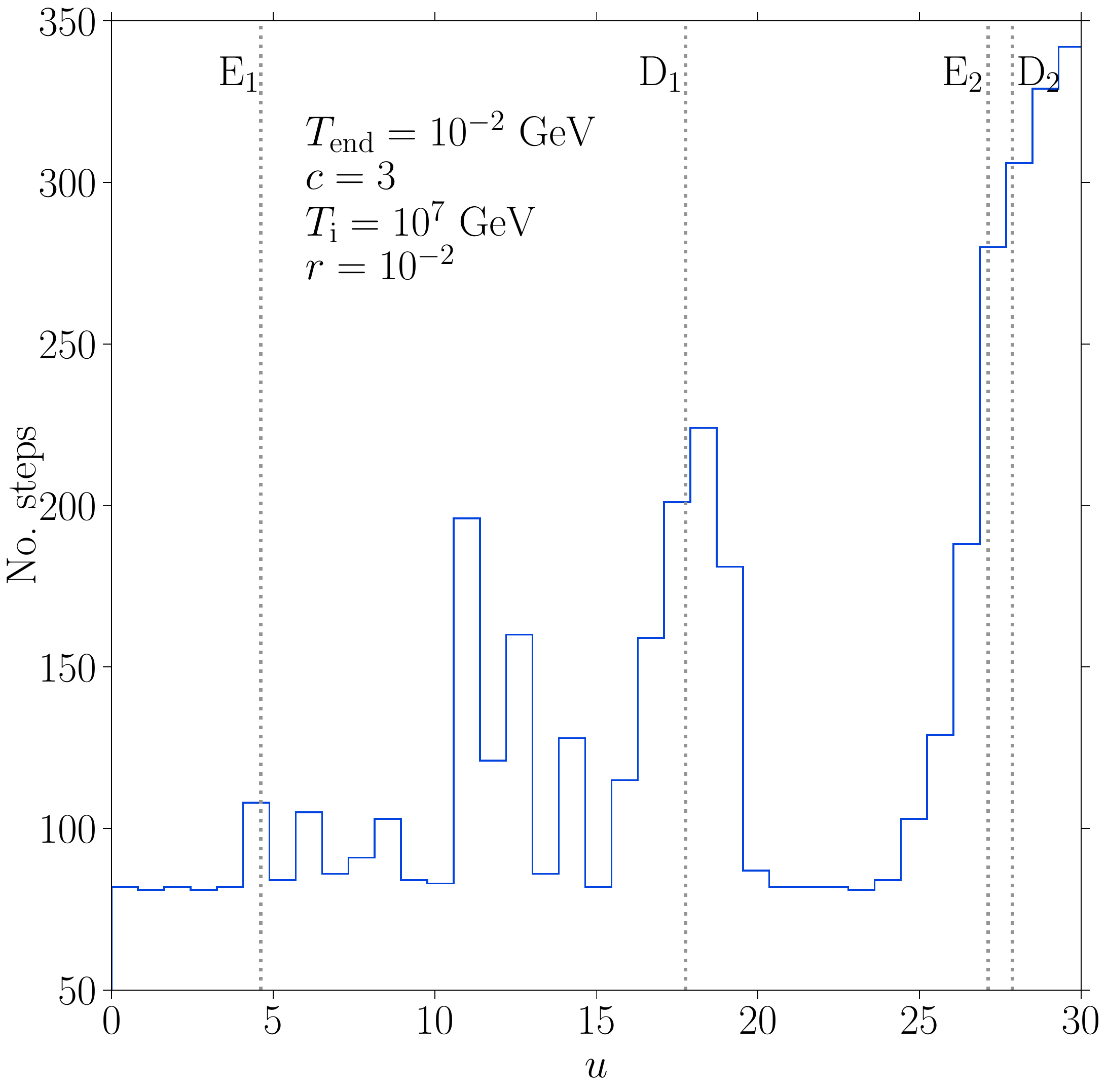}
		\caption{}
		\label{fig:EMD_hist}
	\end{subfigure}
	\caption{			
		({\bfseries a}) The local integration error that corresponds to the temperature of the plasma (black) and $\rhoPhi$ (dashed red) in the case of early matter domination. 
		({\bfseries b}) A histogram of the number of states taken by the solver at various $u$.
		The parameters used are as in Figure~\ref{fig:EMD}.
	}
	\label{fig:error_examples_MD}
\end{figure}
The relative errors associated with the integration of \eqs{eq:dlogfRdu,eq:dlogfPhidu} are shown in \Figs{fig:EMD_err}. As we can see, the local errors appear to be negligible compared to the values of the corresponding quantities, and they appear to need controlling only as long as the decays of $\Phi$ are active. Once the fluid has decayed away, its relative error seems to increase, but this is only because $\rhoPhi$ almost vanishes.
In \Figs{fig:EMD_hist}, we show a histogram of the number of steps taken by the RK solver. In accordance to the increase in the errors, the number of steps increases after $\EI$, which means that the step size changes in order to regulate the integration error.

We also show the relative local errors and histogram corresponding to \Figs{fig:EKD}, which are obtained by changing the parameters of the script to those shown in that figure.   
\begin{figure}[t]
	\begin{subfigure}{0.5\textwidth}
		\includegraphics[width=1\textwidth]{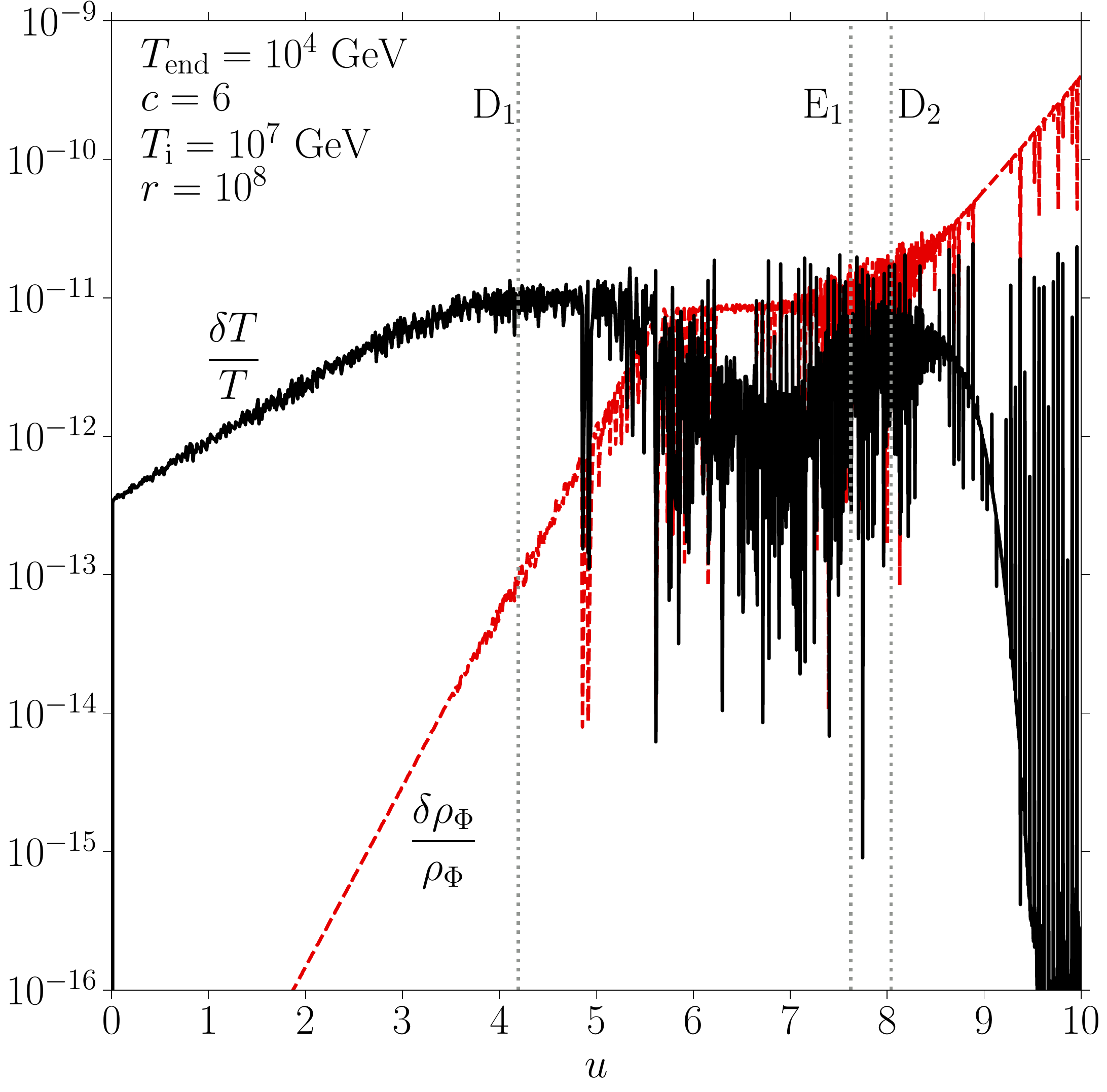}
		\caption{}
		\label{fig:EKD_err}
	\end{subfigure}
	\begin{subfigure}{0.5\textwidth}
		\includegraphics[width=1\textwidth]{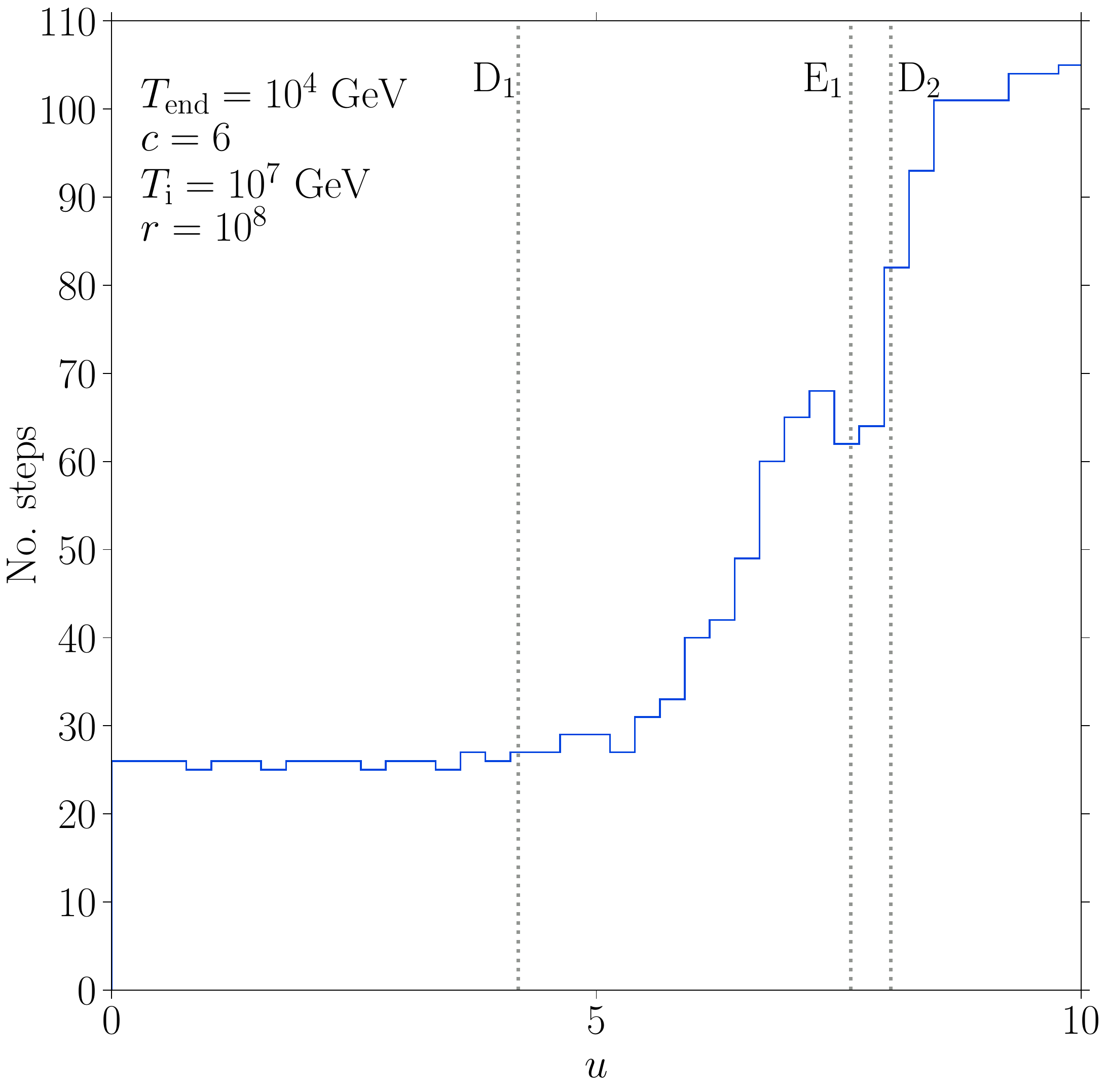}
		\caption{}
		\label{fig:EKD_hist}
	\end{subfigure}
	\caption{Same as  Figure~\ref{fig:error_examples_MD} for the parameters of Figure~\ref{fig:EKD}.}
	\label{fig:error_examples_KD}
\end{figure}
In \Figs{fig:EKD_err,fig:EKD_hist} we can see that the errors are still small, and after $\DI$ the step size is automatically adjusted in order to keep it from increasing.

Figures of errors and histogram such as \Figs{fig:error_examples_MD,fig:error_examples_KD} can be used to identify cases that the solver behaves badly. In general, these plots do not provide the whole picture. It should be considered good practice to use different methods and choices of the various RK parameters (\eg lower of higher values of \pyin{relative_tolerance} and \pyin{absolute_tolerance}) in order to make sure that the results are consistent. Ideally, these results should be also compared against approximate solutions to~\eqs{eq:dlogfRdu,eq:dlogfPhidu} such as the ones explained in detail in ref.~\cite{Arias:2020qty}.

\section{Summary}\label{sec:summary}
We have introduced \nsc, a simple library that simulates the evolution of the Universe assuming the appearance of a decaying fluid. \nsc provides several different methods to solve the radiation-fluid system and also gives access to the local integration errors, which allows the user to try different approaches and decide whether the results can be trusted.

In Section~\refs{sec:equations} we have shown the equations that \nsc solve and briefly described the general form of the solutions we expect. We also showed explicitly the internal notation \nsc uses, which should enable others modify the code in order to meet their needs. 

\nsc can be downloaded and configured following the steps in Section~\refs{sec:first_steps}. This section also broadly describes the various classes and functions that can be used by the user in both \CPP and \PY, with more details given in the Appendix. Complete code examples in \CPP as well as \PY are shown in Section~\refs{sec:example}.

\nsc has limitations as it is assumed that there is only one additional component ($\Phi$) with an equation of state of the given by~\eqs{eq:EOS} and the evolution of its energy density described by~\eqs{eq:drhoPhidt}. \nsc will keep evolving and new versions will be able to handle non-minimal cases such as the ones studied in refs.~\cite{Dienes:2021woi,Ahmed:2021fvt,Barman:2022tzk,Arias:2022qjt}). Another useful extension that will be included in a future version is a module that compares against observational constraints given an inflation model.

\section{Acknowledgements}
This work is supported by the Lancaster–Manchester–Sheffield Consortium for Fundamental Physics, under STFC research grant ST/T001038/1.

\pagebreak
\setcounter{section}{0}
\section*{Appendix}
\appendix

\renewcommand{\theequation}{\Alph{section}.\arabic{equation}}
\setcounter{equation}{0}  
In the appendix, we show a detailed description of the \cppin{Evolution} class in both \CPP and \PY. A detailed description of the \cppin{Cosmo} class as well as a basic introduction of the RK methods used can be found in ref.~\cite{Karamitros:2021nxi}.

\section{\texorpdfstring{\cppin{Evolution}\xspace}~class in \CPP}\label{app:cpp}
\setcounter{equation}{0}
The \cppin{Evolution} class uses the default constructors, so instances are declared trivially as shown in Section~\ref{sec:first_steps}. It has only one public member function, \cppin{solveNSC}, with signature
\begin{cpp}
	template<class LD, const int Solver, class Method>
	bool Evolution<LD,Solver,Method>::solveNSC(const LD &TEND, const LD &c, const LD &Ti, 
							 const LD &ratio, const LD &TSTOP, const LD &umax, Cosmo<LD> *plasma, 
							 const solverArgs<LD> &args={})
\end{cpp}
All the template and required input arguments are summarised in Table~\ref{tab:solveNSC-input}. This function returns \cppin{false} if the integration stops above \cppin{TSTOP} and \cppin{true} otherwise, as a way of providing a simple consistency check. 
The optional argument, \cppin{args}, is an instance of a \cppin{struct} defined as 
\begin{cpp}
 	template<class LD>
	struct solverArgs{
		LD initial_step_size=1e-2, minimum_step_size=1e-8, maximum_step_size=1e-2; 
		LD absolute_tolerance=1e-8, relative_tolerance=1e-8;
		LD beta=0.9, fac_max=1.2, fac_min=0.8;
		unsigned int maximum_No_steps=10000000;
	};
\end{cpp}
This helps us pass to \cppin{solveNSC} only the arguments we wish to change. For example, we can choose {\tt absolute\_tolerance=1e-11} and {\tt beta=0.8} by adding {\tt \{.absolute\_tolerance=1e-11, .beta=0.8\}} in the place of {\tt args}. This is called aggregation in \CPP, and basically allows us to use named arguments. The caveat is that we need to put the arguments in the order thy appear in the definition of \cppin{solverArgs}. For example, using {\tt \{.beta=0.8, .absolute\_tolerance=1e-11\}} will result in a compilation error.

The class \cppin{Evolution} has several member variables that store the results of the solution of the system~\refs{eq:dlogfRdu,eq:dlogfPhidu}:
\begin{enumerate}
	\item \cppin{std::vector<LD> u}: integration steps of $u$.
	\item \cppin{std::vector<LD> T}: values of $T$ (in $\GeV$) that corresponds to every step in \cppin{u}.
	\item \cppin{std::vector<LD> rhoPhi}: values of $\rhoPhi$ (in $\GeV^4$) that corresponds to every step in \cppin{u}.
	\item \cppin{std::vector<LD> dT}: local integration error of $T$ (in $\GeV$) at every \cppin{u}.
	\item \cppin{std::vector<LD> drhoPhi}: local integration error of $\rhoPhi$ (in $\GeV^4$) at every \cppin{u}.
	\item \cppin{LD TE1}: value of $\TEI$ (in $\GeV$). It is initialised at 0. If $\rhoPhi$ never reaches $\rhoR$, This variable will be equal to $\Ti$.
	\item \cppin{LD TE2}: value of $\TEII$ (in $\GeV$). It is initialised at 0. If $\rhoPhi$ never reaches $\rhoR$ or $c>4$, this variable will be equal to $\Ti$. 
	\item \cppin{LD TD1}: value of $\TDI$ (in $\GeV$). It is initialised at 0. If the energy injection rate never becomes more than $10\%$ of the free dilution rate of $\rhoR$, this variable obtains the value of $\Ti$.
	\item \cppin{LD TD2}: value of $\TDII$ (in $\GeV$). It is initialised at 0. If the energy injection rate never becomes more than $10\%$ of the free dilution rate of $\rhoR$, this variable obtains the value of $\Ti$.
	\item \cppin{LD uE1}: value of $\uEI$. It is initialised at 0. If $\rhoPhi$ never reaches $\rhoR$, This variable will be equal to $0$.
	\item \cppin{LD uE2}: value of $\uEII$. It is initialised at 0. If $\rhoPhi$ never reaches $\rhoR$  or $c>4$, this variable will be equal to $0$.
	\item \cppin{LD uD1}: value of $\uDI$. It is initialised at 0. If the energy injection rate never becomes more than $10\%$ of the free dilution rate of $\rhoR$, this variable obtains the value of $0$.
	\item \cppin{LD uD2}: value of $\uDII$. It is initialised at 0. If the energy injection rate never becomes more than $10\%$ of the free dilution rate of $\rhoR$, this variable obtains the value of $0$.	
\end{enumerate} 


\section{\texorpdfstring{\cppin{Evolution}\xspace}~class in \PY }\label{app:py}
\setcounter{equation}{0}
The \pyin{Evolution} class is very similar to the \CPP case. It uses the default constructor, with instances declared as shown in Section~\ref{sec:first_steps}. The \pyin{solveNSC}, has signature
\begin{cpp}
	solveNSC(TEND, c, Ti, ratio, TSTOP, umax, plasma, 
			initial_step_size=1e-2, minimum_step_size=1e-8, maximum_step_size=1e-2, 
			absolute_tolerance=1e-8, relative_tolerance=1e-8, beta=0.9, fac_max=1.2, fac_min=0.8,
			maximum_No_steps=10000000)
\end{cpp}

All the input arguments are summarised in Table~\ref{tab:solveNSC-input}. The difference between this and the \CPP version is that the {\tt plasma} argument is an instance of the \PY \cppin{Cosmo} class. 
This function returns a tuple. The first element of the returned tuple is \pyin{False} if the integration stops above \cppin{TSTOP} and \pyin{True} otherwise, while the second is the time it took to solve the system (in seconds).

The member variables are the same as in \CPP, with {\tt numpy} arrays used in place of the \CPP vectors. However, solving the system using \pyin{solveNSC} does not fill the arrays with the integration steps. The member function
\begin{py}
	getPoints()
\end{py}
Fills the arrays {\tt u}, {\tt T}, and {\tt rhoPhi} with the integration steps.

In order to fill the arrays with the local errors we make use of the member function
\begin{py}
	getErrors()
\end{py}

\section{Quick guide to the user input}\label{app:usr_input}
\setcounter{equation}{0}
In this section, we  present tables with all the inputs the various functions need.
\begin{table}[h!]
	\centering
	\begin{tabular}{l l}
		\hline\\[-0.4cm]
		\multicolumn{2}{c}{\bf User input for solving \eqs{eq:dlogfRdu,eq:dlogfPhidu}.}  \\
		\hline\\[-0.4cm]
		{\tt TEND} & The value of $\Tend$ in $\GeV$.\\
		\hline\\[-0.4cm]
		
		{\tt c} & Value of $c$ as defined through~\eqs{eq:EOS}.\\
		\hline\\[-0.4cm]
		
		{\tt Ti} & The value of $\Ti$ in $\GeV$.\\
		\hline\\[-0.4cm]

		{\tt ratio} & The value of $\rhoPhi/\rhoR$ at $\Ti$.\\
		\hline\\[-0.4cm]

		{\tt plasma} & A pointer to an instance of the {\tt Cosmo} class.\\
		\hline\\[-0.4cm]

		{\tt TSTOP} & Once $T<${\tt TSTOP}, integration stops.\\
		\hline\\[-0.4cm]
		
		{\tt umax } & If $u>${\tt umax} the integration stops.\\
		\hline\\[-0.4cm]
		
		{\tt initial\_stepsize} &  Initial step-size of the solver. Default value: $10^{-2}$.\\ 
		\hline\\[-0.4cm]

		{\tt minimum\_stepsize} & Lower limit of the step-size. Default value:  $10^{-8}$.\\
		\hline\\[-0.4cm]

		{\tt maximum\_stepsize} & Upper limit of the step-size. Default value:  $10^{-2}$.\\
		\hline\\[-0.4cm]

		{\tt absolute\_tolerance} & \multirow{1}{12cm}{Absolute tolerance of the RK solver.  Default value:  $10^{-8}$.}\\\\
		\hline\\[-0.4cm]

		{\tt relative\_tolerance} & \multirow{1}{12cm}{Relative tolerance of the RK solver.  Default value:  $10^{-8}$.}\\\\
		\hline\\[-0.4cm]
		
		{\tt beta} & \multirow{1}{12cm}{Aggressiveness of the adaptation strategy.  Default value:  $0.9$.}\\\\
		\hline\\[-0.4cm]

		{\tt fac\_max}, {\tt fac\_min} &\multirow{1}{12cm}{The step-size does not change more than {\tt fac\_max} and less than {\tt fac\_min} within a trial step . Default values: $1.2$ and $0.8$, respectively.} \\ \\ \\ 
		\hline\\[-0.4cm]
		
		{\tt maximum\_No\_steps} & \multirow{1}{12cm}{If integration needs more than {\tt maximum\_No\_steps} integration stops. Default value: $10^7$.}\\\\
		\hline\\[-0.4cm]
	\end{tabular}
	\caption{Table of the arguments of the \cppin{nsc::Evolution<LD,Solver,Method>::solveNSC} method.}
	\label{tab:solveNSC-input}
\end{table}
\begin{table}[p]
	\centering
	\begin{tabular}{l l}
		
		\hline\\[-0.4cm]
		\multicolumn{2}{c}{\bf User input for interpolating RDOF.}  \\
		\hline\\[-0.4cm]
		
		{\tt path} & \multirow{1}{12cm}{Path to a file with $T$ ($\GeV$), $\heff$, and $\geff$. The temperature has to monotonically increase.}\\\\
		\hline\\[-0.4cm]
		
		{\tt minT} & \multirow{1}{12cm}{The RDOF will be interpolated starting from the closest temperature that satisfies $T_{\rm min}\leq${\tt minT}. Below this, both $\heff$ and $\geff$ will be assumed to be their value at $T_{\rm min}$.}\\\\\\
		\hline\\[-0.4cm]
		
		{\tt maxT} & \multirow{1}{12cm}{The RDOF will be interpolated up to the closest temperature that satisfies $T_{\rm max}\geq${\tt maxT}. Above this, both $\heff$ and $\geff$ will be assumed to be their value at $T_{\rm max}$.}\\\\\\
		\hline\\[-0.4cm]
	\end{tabular}
	\caption{Table of the arguments of the \cppin{nsc::Cosmo<LD>} class constructor.}
	\label{tab:Cosmo-input}
\end{table}
\begin{table}[p]
	\centering
	\begin{tabular}{l l}
		\multicolumn{2}{c}{\bf Template arguments.}  \\
		\hline\\[-0.4cm]
		
		{\tt LD}& \multirow{1}{12cm}{This template argument is the numeric type that \nsc will use. The preferred choice is \cppin{long double}. However, in many cases \cppin{double} can be used. Notice that the instance of the \cppin{Cosmo} class must use identical numeric type to the one that used by the instance of \cppin{Evolution}, in order to be able to run the \cppin{solveNSC} method.}\\\\\\\\\\		
		\hline\\[-0.4cm]
		
		{\tt Solver}& \multirow{1}{12cm}{This is the second template argument of the \cppin{nsc::Evolution<LD,Solver,Method>} class. The available choices are $1$ for Rosenbrock method, and $2$ for explicit RK method.}\\\\\\\\
		\hline\\[-0.4cm]
		
		{\tt Method}& \multirow{1}{12cm}{The third template argument of the \cppin{Evolution} class. Its value depends on the choice of \cppin{Solver}; For {\tt Solver==1}, {\tt Method} can be \cppin{RODASPR2<LD>} (fourth order)~\cite{RangAngermann2005}, \cppin{ROS34PW2<LD>} (third order)~\cite{RANG2015128}, \cppin{GRK4A<LD>} or \cppin{GRK4T<LD>} (fourth order)~\cite{Rentrop1979}. For {\tt Solver==2}, there are a few options, but the only choice that works in most cases is \cppin{DormandPrince<LD>} (fourth order)~\cite{DORMAND198019}. The {\tt Method} classes need a template argument, \cppin{LD}, which must be the same as the first template argument of the \cppin{nsc::Evolution<LD,Solver,Method>} class. If one defines their own Butcher table, they would have to follow their definitions and assumptions.}\\\\\\\\\\\\\\\\\\\\
		\hline
	\end{tabular}
	\caption{Template arguments of the various \nsc classes.}
	\label{tab:template-arguments}
\end{table}
\begin{table}[p]
	\centering
	\begin{tabular}{l l}
		\multicolumn{2}{c}{\bf User compile-time options. Variables in the various {\tt Definitions.mk} files.}  \\
		\hline\\[-0.4cm]

		{\tt rootDir}& \multirow{1}{12cm}{The relative path of root directory of \nsc. Relevant only when compiling using {\tt make}. Available in all {\tt Definitions.mk}.}\\\\		
		\hline\\[-0.4cm]
		
		{\tt LONG}& \multirow{1}{12cm}{{\tt long} for \cppin{long double} or empty for \cppin{double}. This is defines a macro in the source files of the various \CPP examples. Available in {\tt Definitions.mk} inside the various subdirectories of {\tt NSCpp/UserSpace/Cpp}.}\\\\\\\\		
		\hline\\[-0.4cm]

		{\tt LONGpy}& \multirow{1}{12cm}{{\tt long} or empty. Same as {\tt LONG}, applies to the \PY modules. Available in {\tt NSCpp/Definitions.mk}.}\\\\		
		\hline\\[-0.4cm]

		{\tt SOLVER}& \multirow{1}{12cm}{In order to use a Rosenbrock method {\tt SOLVER}=$1$. For explicit RK method, {\tt SOLVER}=$2$. This defines a macro that is passes as the second template argument of \cppin{nsc::Evolution<LD,Solver,Method>}.  The corresponding variable in {\tt NSCpp/Definitions.mk} applies to the \PY modules. The variable in {\tt NSCpp/UserSpace/Cpp/NSC/Definitions.mk} applies to the example in the same directory.}\\\\\\\\\\\\\\		
		\hline\\[-0.4cm]

		{\tt METHOD}& \multirow{1}{12cm}{Depending on the solver, this variable should name one of its available methods. For {\tt SOLVER}=$1$, {\tt METHOD}={\tt RODASPR2}(fourth order) or {\tt ROS34PW2}(third order). For {\tt SOLVER}=$2$, {\tt METHOD}={\tt DormandPrince }(seventh order). There is a macro ({\tt METHOD}) used by the shared library {\tt NSCpp/lib/libNSC.so}. The corresponding variable in {\tt NSCpp/Definitions.mk} applies to the \PY modules. The variable in {\tt NSCpp/UserSpace/Cpp/NSC/Definitions.mk} applies to the example in that directory.}\\\\\\\\\\\\\\\\
		 		
		\hline\\[-0.4cm]
		
		\multicolumn{2}{c}{\bf Compiler options}  \\
		\hline\\[-0.4cm]
		
		{\tt CC} &  \multirow{1}{12cm}{The preferred \CPP compiler ({\tt g++} by default). Corresponding variable in all {\tt Definitions.mk} files.} \\\\
		\hline\\[-0.4cm]
		
		{\tt OPT} &  \multirow{1}{12cm}{Available options are {\tt OPT}={\tt O1}, {\tt O2}, {\tt O3} (default). This variable defines the optimization level of the compiler. The variable can be changed in all {\tt Definitions.mk} files. In the root directory of \nsc, the optimization level applies to the python modules (\ie the shared libraries), while in the subdirectories of {\tt NSCpp/UserSpace/Cpp} it only applies to example inside them.}   \\\\\\\\\\\\
		\hline\\[-0.4cm]

	\end{tabular}
	\caption{User compile-time input and options. These are available in the various {\tt Definitions.mk} files, which are used when compiling using {\tt make}.}
	\label{tab:compile_time-options}
\end{table}

\pagebreak
\bibliography{refs}{}

\providecommand{\href}[2]{#2}\begingroup\raggedright\begin{thebibliography}{10}

\bibitem{McDonald:1989jd}
J.~McDonald, {\it {{WIMP} Densities in Decaying Particle Dominated Cosmology}},
   {\em Phys. Rev. D} {\bf 43} (1991) 1063--1068.

\bibitem{DEramo:2017gpl}
F.~D'Eramo, N.~Fernandez, and S.~Profumo, {\it {When the Universe Expands Too
  Fast: Relentless Dark Matter}},  {\em JCAP} {\bf 05} (2017) 012,
  [\href{http://arxiv.org/abs/1703.04793}{{\tt arXiv:1703.04793}}].

\bibitem{Redmond:2017tja}
K.~Redmond and A.~L. Erickcek, {\it {New Constraints on Dark Matter Production
  during Kination}},  {\em Phys. Rev. D} {\bf 96} (2017), no.~4 043511,
  [\href{http://arxiv.org/abs/1704.01056}{{\tt arXiv:1704.01056}}].

\bibitem{DEramo:2017ecx}
F.~D'Eramo, N.~Fernandez, and S.~Profumo, {\it {Dark Matter Freeze-in
  Production in Fast-Expanding Universes}},  {\em JCAP} {\bf 02} (2018) 046,
  [\href{http://arxiv.org/abs/1712.07453}{{\tt arXiv:1712.07453}}].

\bibitem{Bernal:2020bfj}
N.~Bernal, J.~Rubio, and H.~Veerm\"ae, {\it {Boosting Ultraviolet Freeze-in in
  NO Models}},  {\em JCAP} {\bf 06} (2020) 047,
  [\href{http://arxiv.org/abs/2004.13706}{{\tt arXiv:2004.13706}}].

\bibitem{Arias:2020qty}
P.~Arias, D.~Karamitros, and L.~Roszkowski, {\it {Frozen-in fermionic singlet
  dark matter in non-standard cosmology with a decaying fluid}},  {\em JCAP}
  {\bf 05} (2021) 041, [\href{http://arxiv.org/abs/2012.07202}{{\tt
  arXiv:2012.07202}}].

\bibitem{Arias:2021rer}
P.~Arias, N.~Bernal, D.~Karamitros, C.~Maldonado, L.~Roszkowski, and
  M.~Venegas, {\it {New opportunities for axion dark matter searches in
  nonstandard cosmological models}},  {\em JCAP} {\bf 11} (2021) 003,
  [\href{http://arxiv.org/abs/2107.13588}{{\tt arXiv:2107.13588}}].

\bibitem{Barman:2021ifu}
B.~Barman, P.~Ghosh, F.~S. Queiroz, and A.~K. Saha, {\it {Scalar multiplet dark
  matter in a fast expanding Universe: Resurrection of the desert region}},
  {\em Phys. Rev. D} {\bf 104} (2021), no.~1 015040,
  [\href{http://arxiv.org/abs/2101.10175}{{\tt arXiv:2101.10175}}].

\bibitem{Dienes:2021woi}
K.~R. Dienes, L.~Heurtier, F.~Huang, D.~Kim, T.~M.~P. Tait, and B.~Thomas, {\it
  {Stasis in an expanding universe: A recipe for stable mixed-component
  cosmological eras}},  {\em Phys. Rev. D} {\bf 105} (2022), no.~2 023530,
  [\href{http://arxiv.org/abs/2111.04753}{{\tt arXiv:2111.04753}}].

\bibitem{Banerjee:2022fiw}
A.~Banerjee and D.~Chowdhury, {\it {Fingerprints of freeze-in dark matter in an
  early matter-dominated era}},  {\em SciPost Phys.} {\bf 13} (2022), no.~2
  022, [\href{http://arxiv.org/abs/2204.03670}{{\tt arXiv:2204.03670}}].

\bibitem{Hardy:2018bph}
E.~Hardy, {\it {Higgs portal dark matter in non-standard cosmological
  histories}},  {\em JHEP} {\bf 06} (2018) 043,
  [\href{http://arxiv.org/abs/1804.06783}{{\tt arXiv:1804.06783}}].

\bibitem{Bernal:2018kcw}
N.~Bernal, C.~Cosme, T.~Tenkanen, and V.~Vaskonen, {\it {Scalar singlet dark
  matter in non-standard cosmologies}},  {\em Eur. Phys. J. C} {\bf 79} (2019),
  no.~1 30, [\href{http://arxiv.org/abs/1806.11122}{{\tt arXiv:1806.11122}}].

\bibitem{Arias:2019uol}
P.~Arias, N.~Bernal, A.~Herrera, and C.~Maldonado, {\it {Reconstructing
  Non-standard Cosmologies with Dark Matter}},  {\em JCAP} {\bf 10} (2019) 047,
  [\href{http://arxiv.org/abs/1906.04183}{{\tt arXiv:1906.04183}}].

\bibitem{Allahverdi:2019jsc}
R.~Allahverdi and J.~K. Osi\'nski, {\it {Freeze-in Production of Dark Matter
  Prior to Early Matter Domination}},  {\em Phys. Rev. D} {\bf 101} (2020),
  no.~6 063503, [\href{http://arxiv.org/abs/1909.01457}{{\tt
  arXiv:1909.01457}}].

\bibitem{Bernal:2019mhf}
N.~Bernal, F.~Elahi, C.~Maldonado, and J.~Unwin, {\it {Ultraviolet Freeze-in
  and Non-Standard Cosmologies}},  {\em JCAP} {\bf 11} (2019) 026,
  [\href{http://arxiv.org/abs/1909.07992}{{\tt arXiv:1909.07992}}].

\bibitem{Cosme:2020mck}
C.~Cosme, M.~Dutra, T.~Ma, Y.~Wu, and L.~Yang, {\it {Neutrino portal to FIMP
  dark matter with an early matter era}},  {\em JHEP} {\bf 03} (2021) 026,
  [\href{http://arxiv.org/abs/2003.01723}{{\tt arXiv:2003.01723}}].

\bibitem{Vilenkin:1982wt}
A.~Vilenkin and L.~H. Ford, {\it {Gravitational Effects upon Cosmological Phase
  Transitions}},  {\em Phys. Rev. D} {\bf 26} (1982) 1231.

\bibitem{Coughlan:1983ci}
G.~D. Coughlan, W.~Fischler, E.~W. Kolb, S.~Raby, and G.~G. Ross, {\it
  {Cosmological Problems for the Polonyi Potential}},  {\em Phys. Lett. B} {\bf
  131} (1983) 59--64.

\bibitem{Ratra:1987rm}
B.~Ratra and P.~J.~E. Peebles, {\it {Cosmological Consequences of a Rolling
  Homogeneous Scalar Field}},  {\em Phys. Rev. D} {\bf 37} (1988) 3406.

\bibitem{Giudice:2000ex}
G.~F. Giudice, E.~W. Kolb, and A.~Riotto, {\it {Largest temperature of the
  radiation era and its cosmological implications}},  {\em Phys. Rev. D} {\bf
  64} (2001) 023508, [\href{http://arxiv.org/abs/hep-ph/0005123}{{\tt
  hep-ph/0005123}}].

\bibitem{Gardner:2004in}
C.~L. Gardner, {\it {Quintessence and the transition to an accelerating
  universe}},  {\em Nucl. Phys. B} {\bf 707} (2005) 278--300,
  [\href{http://arxiv.org/abs/astro-ph/0407604}{{\tt astro-ph/0407604}}].

\bibitem{Dalianis:2018afb}
I.~Dalianis and Y.~Watanabe, {\it {Probing the BSM physics with CMB precision
  cosmology: an application to supersymmetry}},  {\em JHEP} {\bf 02} (2018)
  118, [\href{http://arxiv.org/abs/1801.05736}{{\tt arXiv:1801.05736}}].

\bibitem{Tsujikawa:2013fta}
S.~Tsujikawa, {\it {Quintessence: A Review}},  {\em Class. Quant. Grav.} {\bf
  30} (2013) 214003, [\href{http://arxiv.org/abs/1304.1961}{{\tt
  arXiv:1304.1961}}].

\bibitem{Allahverdi:2020bys}
R.~Allahverdi et~al., {\it {The First Three Seconds: a Review of Possible
  Expansion Histories of the Early Universe}},
  \href{http://arxiv.org/abs/2006.16182}{{\tt arXiv:2006.16182}}.

\bibitem{Karamitros:2021nxi}
D.~Karamitros, {\it {MiMeS: Misalignment mechanism solver}},  {\em Comput.
  Phys. Commun.} {\bf 275} (2022) 108311,
  [\href{http://arxiv.org/abs/2110.12253}{{\tt arXiv:2110.12253}}].

\bibitem{Dutra:2021phm}
M.~Dutra and Y.~Wu, {\it {EvoEMD: cosmic evolution with an early
  matter-dominated era}},  \href{http://arxiv.org/abs/2111.15665}{{\tt
  arXiv:2111.15665}}.

\bibitem{NaBBODES}
D.~Karamitros, {\it {\href{https://github.com/dkaramit/NaBBODES}{NaBBODES}: Not
  a Black Box Ordinary Differential Equation Solver in {C++}}},  2019.

\bibitem{SimpleSplines}
D.~Karamitros, {\it
  {\href{https://github.com/dkaramit/SimpleSplines}{SimpleSplines}: A
  header-only library for linear and cubic spline interpolation in {C++}}},
  2021.

\bibitem{Saikawa:2020swg}
K.~Saikawa and S.~Shirai, {\it {Precise WIMP Dark Matter Abundance and Standard
  Model Thermodynamics}},  {\em JCAP} {\bf 08} (2020) 011,
  [\href{http://arxiv.org/abs/2005.03544}{{\tt arXiv:2005.03544}}].

\bibitem{Planck:2018jri}
{\bf Planck} Collaboration, Y.~Akrami et~al., {\it {Planck 2018 results. X.
  Constraints on inflation}},  {\em Astron. Astrophys.} {\bf 641} (2020) A10,
  [\href{http://arxiv.org/abs/1807.06211}{{\tt arXiv:1807.06211}}].

\bibitem{harris2020array}
C.~R. Harris, K.~J. Millman, S.~J. van~der Walt, R.~Gommers, P.~Virtanen,
  D.~Cournapeau, E.~Wieser, J.~Taylor, S.~Berg, N.~J. Smith, R.~Kern, M.~Picus,
  S.~Hoyer, M.~H. van Kerkwijk, M.~Brett, A.~Haldane, J.~F. del R{\'{i}}o,
  M.~Wiebe, P.~Peterson, P.~G{\'{e}}rard-Marchant, K.~Sheppard, T.~Reddy,
  W.~Weckesser, H.~Abbasi, C.~Gohlke, and T.~E. Oliphant, {\it Array
  programming with {NumPy}},  {\em Nature} {\bf 585} (Sept., 2020) 357--362.

\bibitem{RangAngermann2005}
J.~Rang and L.~Angermann, {\it New rosenbrock w-methods of order 3 for partial
  differential algebraic equations of index 1},  {\em BIT Numerical
  Mathematics} {\bf 45} (2005) 761–787.

\bibitem{Ahmed:2021fvt}
A.~Ahmed, B.~Grzadkowski, and A.~Socha, {\it {Implications of time-dependent
  inflaton decay on reheating and dark matter production}},  {\em Phys. Lett.
  B} {\bf 831} (2022) 137201, [\href{http://arxiv.org/abs/2111.06065}{{\tt
  arXiv:2111.06065}}].

\bibitem{Barman:2022tzk}
B.~Barman, N.~Bernal, Y.~Xu, and O.~Zapata, {\it {Ultraviolet freeze-in with a
  time-dependent inflaton decay}},  {\em JCAP} {\bf 07} (2022), no.~07 019,
  [\href{http://arxiv.org/abs/2202.12906}{{\tt arXiv:2202.12906}}].

\bibitem{Arias:2022qjt}
P.~Arias, N.~Bernal, J.~K. Osi\'nski, and L.~Roszkowski, {\it {Dark Matter
  Axions in the Early Universe with a Period of Increasing Temperature}},
  \href{http://arxiv.org/abs/2207.07677}{{\tt arXiv:2207.07677}}.

\bibitem{RANG2015128}
J.~Rang, {\it Improved traditional rosenbrock–wanner methods for stiff odes
  and daes},  {\em Journal of Computational and Applied Mathematics} {\bf 286}
  (2015) 128--144.

\bibitem{Rentrop1979}
P.~Rentrop and P.~Kaps, {\it {Generalized Runge-Kutta Methods of Order Four
  with Stepsize Control for Stiff Ordinary Differential Equations.}},  {\em
  Numerische Mathematik} {\bf 33} (1979) 55--68.

\bibitem{DORMAND198019}
J.~Dormand and P.~Prince, {\it A family of embedded runge-kutta formulae},
  {\em Journal of Computational and Applied Mathematics} {\bf 6} (1980), no.~1
  19--26.

\end{thebibliography}\endgroup
\bibliographystyle{JHEP}                        

\end{document}